\documentclass[prl,aps,twocolumn,showpacs,bibnotes,epsf]{revtex4}
\usepackage{graphicx, verbatim}
\usepackage{amsmath}
\newcommand{\Journal}[4]{#1 \textbf{#2}, #3 (#4)}
\usepackage{epstopdf}
\usepackage{CJK}
\usepackage[utf8]{inputenc}
\usepackage[english]{babel}
\usepackage{hyperref}
\hypersetup{colorlinks=true,linkcolor=blue,filecolor=magenta,urlcolor=cyan}

\begin{document}
\begin{CJK*}{GB}{gbsn}
\title{Dynamical mode coupling and coherence in spin Hall nano-oscillator with perpendicular magnetic anisotropy}
\author{Lina Chen$^1$}
\author{S. Urazhdin$^2$}
\author{Y.W. Du$^{1}$}
\author{R. H. Liu$^{1,2}$}
\affiliation{
$^1$National Laboratory of Solid State Microstructures, School of Physics and Collaborative Innovation \\Center of Advanced Microstructures, Nanjing University, Nanjing 210093, China\\
$^2$Department of Physics, Emory University, Atlanta, GA 30322, USA.}

\begin{abstract}
We experimentally study the dynamical modes excited by spin current in Spin Hall nano-oscillators based on the Pt/[Co/Ni] multilayers with perpendicular magnetic anisotropy. Both propagating spin wave and localized solitonic modes of the oscillation are achieved and controlled by varying the applied magnetic field and current. At room temperature, the generation linewidth broadening associated with mode hopping was observed at currents close to the transition between different modes and in the mode coexistence regimes. The mode hopping was suppressed at cryogenic temperatures, confirming that the coupling between modes is mediated by thermal magnons. We also demonstrate that coherent single-mode oscillations with linewidth of 5 MHz can be achieved without applying external magnetic field. Our results provide insight into the mechanisms controlling the dynamical coherence in nanomagnetic oscillators, and guidance for optimizing their applications in spin wave-based electronics.

\end{abstract}

\pacs{75.78.-n, 75.75.-c, 75.30.Ds}

\maketitle
\end{CJK*}

Spin transfer torque (STT), a torque exerted on the magnetization by the injected spin current, can counteract the natural dynamical damping in magnetic systems, resulting in magnetization reversal or sustained magnetization precession~\cite{slon1,berger}. Spin torque nano-oscillators (STNO), magnetic nanodevices based on the latter effect, can serve as nanoscale sources of microwave signals and spin waves, with possible applications in rf electronics, spin wave-based electronic (magnonic) devices~\cite{magnonics}, and neuromorphic computing~\cite{TorrejonNature2017}.

One of the main shortcomings of STNOs is their large microwave generation linewidth, motivating the ongoing experimental and theoretical studies of the mechanisms controlling the coherence of the dynamical magnetization states induced by STT. A theory based on the single-mode nonlinear oscillator approximation, developed by Slavin and co-workers, predicted that the dynamical coherence is determined by thermal fluctuations enhanced by the dynamical nonlinearity~\cite{nonlineartheory1,nonlineartheory2}. However, real nanomagnetic systems are characterized by quasi-continuous spectrum of dynamical modes. As a result, mode coexistence, mode hopping, and/or periodic mode transitions are commonly observed in STNOs~\cite{temline,Fert-Tem-Vor,thermaleffect,SWsCoex,hoppingmech,modecoupling}, significantly affecting the dynamical coherence.

In mode hopping, the magnetic system experiences random transitions between different dynamical states, with only a single mode excited at any given instant of time, indicating that the modes compete with each other and are generally incompatible. It is also possible for different modes to simultaneously coexist, which can be facilitated by the spatial separation of different modes, due, for example, to the current-induced Oersted field\cite{SWsCoex}, the spatially inhomogeneous dipolar fields, and/or the spatial inhomogeneity of the magnetic properties of the system. To account for these observed effects, theories accounting for the multimodal behaviors of STNO have been recently developed~\cite{modehoppingtheory,modecoupling2}. They provide insight into the mode coupling mechanisms essential for developing efficient microwave and spin-wave applications of STNOs. However, experimental verification of the mode coupling mechanisms for some of the most actively researched STNO types described below is still lacking.


\begin{figure}[htbp]
\centering
\includegraphics[width=0.45\textwidth]{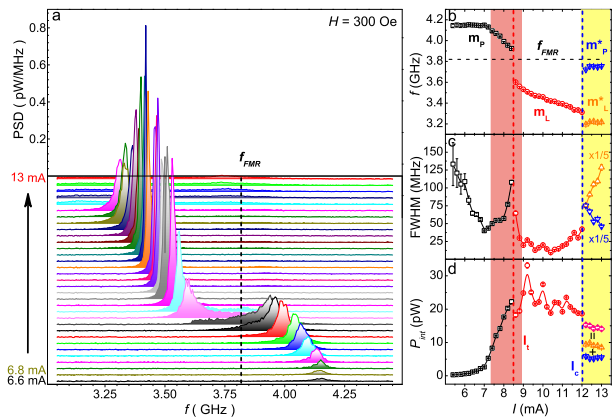}
\caption{Dependence of the microwave generation characteristics of SHNO on current, at $H = 300$~Oe and $T = 295$~K. (a) Generation spectra, at current $I$ varied between 6.6 mA and 13 mA in 0.2 mA steps. The dashed line is the ferromagnetic resonance (FMR) frequency $f_{FMR}$ determined by the spin torque FMR (ST-FMR) technique. (b)-(d) Current dependence of the central generation frequency (b), full width at half maximum (FWHM) (c), and integral intensity (d) of the three dominant modes $m_P$, $m_L$ and $m^*_P$ defined in the text. The mode transition current $I_t$ and the onset current $I_C$ of mode coexistence are marked by the vertical dashed lines. The two shadowed regions mark the mode hopping (at 7.3~mA$<$$I<8.9$~mA) and mode coexistence (at $I>12$~mA) regimes. The spectral characteristics in (b)-(d) were extracted from the multipeak Lorentzian fitting of the spectra. At $I>12$~mA, the FWHMs of both $m^*_P$ and $m_L$ modes in panel (c) are divided by 5, as marked by $\times 1/5$. The symbols ($+$, $\parallel$) in panel (d) indicate the sum of the integral intensities $P_{int}$ of $m^*_P$ and $m_L$ modes, giving the total intensity shown with diamonds.
}\label{fig1}
\end{figure}

Traditionally, STT applications utilized magnetic multilayers, where the current exerting STT on the "free" magnetic layer was spin-polarized by a separate spin-polarizing ferromagnet~\cite{tsoiprl,cornellorig}. More recently,
an alternative, simpler and potentially more efficient approach has emerged~\cite{miron,liulq,liuprb,Demidov_SHO,liuSHNO1}, relying on spin currents produced by the spin Hall effect (SHE)~\cite{Diakonov}, or on the spin-orbit torque (SOT)~\cite{manchon} exerted on the magnetic interface due to the Rashba effect~\cite{rashba}. These advancements have enabled the development of a novel type of STNO - the spin Hall nano-oscillator (SHNO) comprising a bilayer of an efficient spin Hall material and a ferromagnet (FM), without the need for a separate spin-polarizer~\cite{Demidov_SHO,liuSHNO1}.

In SHNO, pure spin current generated by SHE in material with a large SHE, such as Pt and W, is locally injected into the adjacent FM layer, exciting either localized magnetization dynamics or propagating spin waves~\cite{NSHNO,NWSHNO,nanoSHNO2,secondpeak,rhliuPRL2015,mazraati,mutualSHNO1,mutualSHNO2}. The potential advantages of SHNO, compared to the traditional multilayer STNOs, include simple planar structure enabling simultaneous fabrication and synchronization of multiple nano-oscillators~\cite{mutualSHNO1}, and their straightforward incorporation into magnonic structures~\cite{PRSHNO}. Additionally, since SHNO do not require charge currents to flow through the magnetic layers, they are compatible with low-loss insulating magnetic materials, enabling improved efficiency and reduced Joule heating~\cite{SHNOYIG}. Despite intense research of SHNO, the nature of transitions among different dynamical modes in SHNO, and their effects on the dynamical coherence, remain largely unexplored.

Here, we report an experimental study of the microwave generation characteristics of SHNO based on the FM film with perpendicular magnetic anisotropy (PMA). We analyze the dependence on the applied magnetic field, excitation current, and temperature, providing insight into the  interactions among the dynamical modes, the effects of mode hopping and mode coexistence on the dynamical coherence and linewidth broadening of SHNO. Additionally, we demonstrate the possibility to achieve single-mode dynamics in the absence of applied magnetic field. Our experimental results provide  a valuable test for the proposed theories of multimode dynamics, and  suggest avenues for the control of spectral characteristics and generation power of SHNO for device applications.

\begin{figure*}[htbp]
\centering
\includegraphics[width=0.7\textwidth]{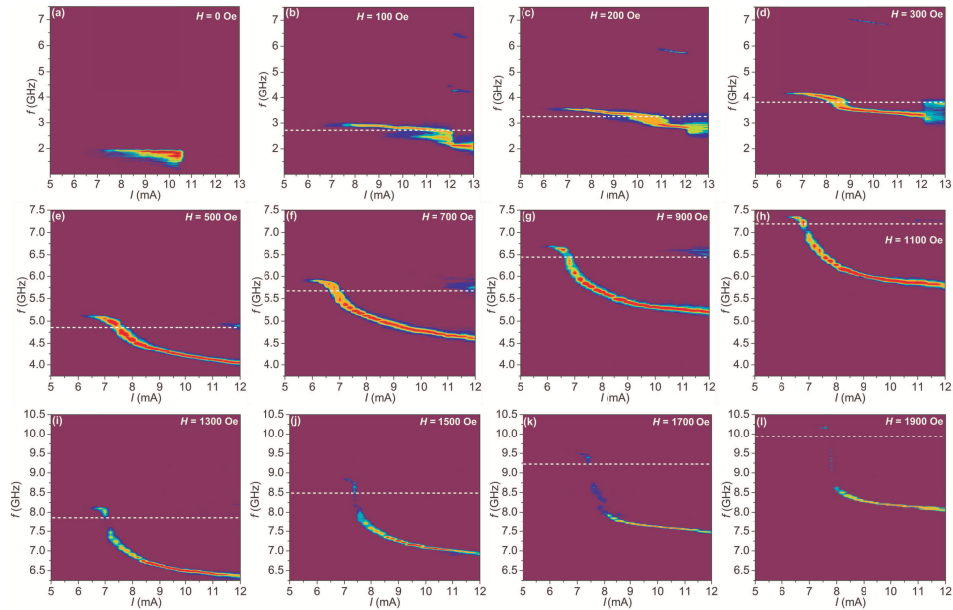}
\caption{Pseudocolor maps of the dependence of the generated microwave spectra on current at $T=295$~ K, at the magnetic field $H$ = 0 Oe (a), 100 Oe (b), 200 Oe (c), 300 Oe (d), 500 Oe (e), 700 Oe (f), 900 Oe (g), 1100 Oe (h), 1300 Oe (i), 1500 Oe (j), 1700 Oe (k) and 1900 Oe (l). Dashed horizontal lines show the ferromagnetic resonance frequency $f_{FMR}$ of the device determined by the ST-FMR technique.}\label{fig2}
\end{figure*}

The studied device is based on a Pt(5)/[Co(0.2)/Ni(0.3)]$_6$/SiO$_2$(3) magnetic multilayer, deposited on the sapphire substrate by magnetron sputtering at room temperature. Thicknesses are given in nanometers. The magnetic properties of the multilayer were characterized by measurements of the anomalous Hall effect (AHE), the anisotropic magnetoresistance (AMR), and magneto-optical Kerr effect (MOKE). The magnetic multilayer film has a well-defined perpendicular magnetic anisotropy (PMA). Magnetic hysteresis loop measurements indicated the presence of bubble magnetic domains when a small field was applied in the plane of the film. The SHNO device consisted of the Pt/[Co/Ni] multilayer patterned into a disk with diameter of 4~$\mu$m, with two pointed Au(100) electrodes, separated by a $\sim$100~nm gap, fabricated on top of the disk, similar to the previously studied planar point contact PMA SHNO~\cite{rhliuPRL2015}. The microwave signals detected in our measurements were generated due to the AMR of the [Co/Ni] film, whose magnetization experienced precession induced by the spin current injected by the Pt layer. To enable spin current-induced dynamics and microwave generation, the spectroscopic measurements were performed with $H$ tilted by $5^\circ$ relative to the film plane,  forming an angle $\theta=60^\circ$ relative to the current direction.

\begin{figure}[htbp]
\centering
\includegraphics[width=0.45\textwidth]{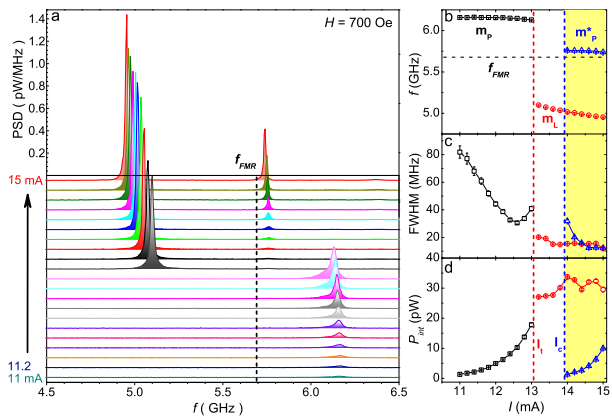}
\caption{Dependence of the microwave generation characteristics on current, at $H = 700$~Oe and $T = 6$~K. (a) Generation spectra, at current $I$ varied between 11 mA and 15 mA in 0.2 mA steps. The dashed line is $f_{FMR}$ determined by the ST-FMR technique. (b)-(d) Current dependence of the central generation frequency (b), FWHM (c), and integral intensity (d) of the dominant modes labeled $m_P$, $m_L$ and $m^*_P$. The mode transition current $I_t$ and the onset current $I_C$ of two-mode coexistence are marked by the vertical dashed lines. The shadowed region marks the mode coexistence regime. The spectral characteristics in (b)-(d) were extracted from the multipeak Lorentzian fitting of the spectra.
}\label{fig3}
\end{figure}

\begin{figure*}[htbp]
\centering
\includegraphics[width=0.7\textwidth]{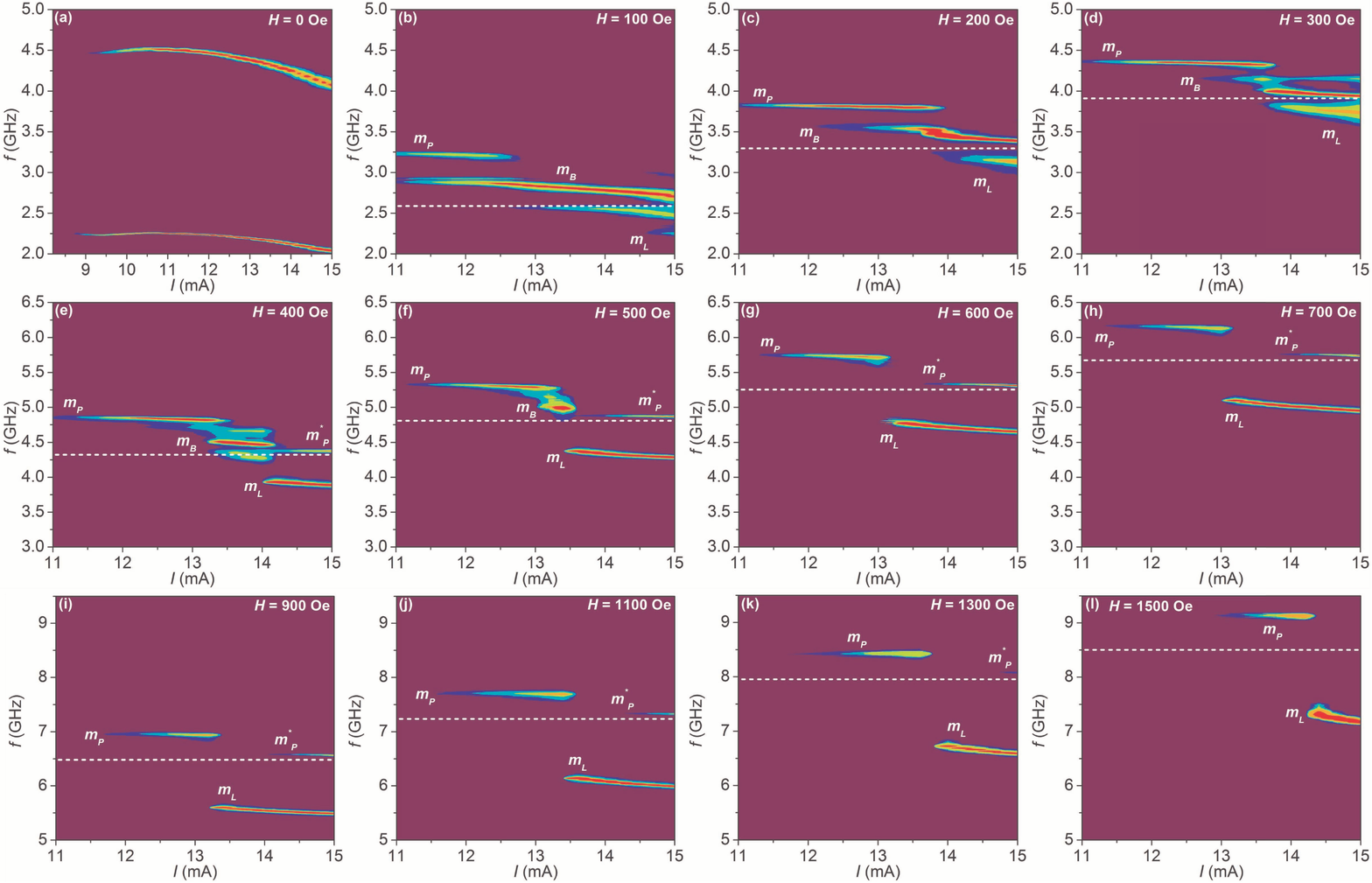}
\caption{Pseudocolor maps of the dependence of the generated microwave spectra on current at $T=6$~ K, at the magnetic field $H$ = 0 Oe (a), 100 Oe (b), 200 Oe (c), 300 Oe (d), 400 Oe (e), 500 Oe (f), 600 Oe (g), 700 Oe (h), 900 Oe (i), 1100 Oe (j), 1300 Oe (k) and 1500 Oe (l). Dashed horizontal lines show $f_{FMR}$ determined by the ST-FMR technique. $m_P$, $m_L$ and $m^*_P$ mark the three distinct modes defined in the text.}\label{fig4}
\end{figure*}


At room temperature $T$=295~K, magnetization dynamics was observed above the auto-oscillation onset current $I_{on}$, at magnetic fields $H$ ranging from 0 to almost $1.2$~kOe, as illustrated in Figs.~\ref{fig1},~\ref{fig2}. Figure~\ref{fig1} shows a representative dependence of the spectral characteristics on current $I$, obtained at $H = 300$~Oe.  At the onset current $I_{on} = 5.4$~mA, the oscillation frequency $f$ = 4.15~GHz was above the ferromagnetic resonance frequency $f_{FMR}=3.82$~GHz determined by the spin torque ferromagnetic resonance (ST-FMR) technique [Fig.~\ref{fig1}(a)]~\cite{mosendz}. These characteristics are consistent with Slonczewski's quasi-linear theory of propagating spin wave emission by STNO with PMA free layer~\cite{slon1}. The oscillation exhibited a weak blueshift and a linear decrease of linewidth with increasing $I$. The linewidth decreased to a minimum value of 50 MHz at the same current $I_{p1}$ = 7.0~mA as the maximum peak intensity.

At currents  above $I$ = 7.2 mA, the frequency started to rapidly decrease, accompanied by an increase of the linewidth and the generated power. At a current $I_t$ = 8.5 mA, the oscillation frequency abruptly dropped to $f=3.6$~GHz, below $f_{FMR}=3.82$~GHz, and the linewidth also abruptly decreased. These behaviors can be attributed to the transition to a different auto-oscillation mode that does not belong to the linear spin-wave spectrum. In contrast to the propagating spin wave mode at $I<I_t$, in this regime the oscillation exhibited a red shift with increasing current. Similar spectral features were previously observed in SHNO with in-plane magnetic anisotropy~\cite{Demidov_SHO,liuSHNO1} and the conventional multilayer STNO~\cite{SWsCoex}, and were identified with the nonlinear self-localized spin wave "bullet" mode~\cite{nonlineartheory1}. With a further increase of current, an additional broad low-intensity peak emerges in the spectrum above $I_c=12$~mA, at frequency slightly below $f_{FMR}$. This is correlated with the decrease of intensity and broadening of the low-frequency "bullet" mode. The new peak can be attributed to the quasi-propagating mode that becomes weakly localized due to the Oersted field of the driving current, which is directed almost opposite to the applied field $H$~\cite{rhliuPRL2015}.

Multimodal dynamics such as observed in Fig.~\ref{fig1} near $I_t$ and above $I_c$, is generally associated either with mode hopping or mode coexistence. The former is characterized by random transitions between mutually incompatible modes. Based on the abrupt crossover between qualitatively different dynamical behaviors at $I_t=8.5$~mA, we can attribute this transition to mode hopping. Meanwhile, the behaviors at $I>12$~mA can be attributed to the emergence of mode coexistence. We provide further evidence for this interpretation below. We note that the linewidth doubles due to mode hopping near $I_t$, and increases by more than a factor of 5 due to the mutilmode excitation above $I_C$, demonstrating the importance of understanding the underlying mechanisms for the ability to control the spectral characteristics of SHNO.

Since the magnetic configuration of the system depends on both the driving current and the applied magnetic field $H$, further insight into the nature of the dynamical magnetization states is provided by analyzing the effects of varying $H$ [Fig.~\ref{fig2}]. At small fields $H<100$~Oe, we observed a hysteresis of spectral characteristics consistent with magnetic history-dependent pinning of the magnetic bubbles on imperfections. In constrast to STNO with in-plane anisotropy, the studied SHNO device with PMA generated microwave signals even in the absence of applied field, with an almost current-independent frequency $f$ = 1.8~GHz, at currents above $6.4$~mA [Fig.~\ref{fig1}(a)]. The spectrum exhibited an asymmetric lineshape with a long low frequency tail, which is likely associated with the interplay between the current-induced dynamics and thermal hopping of magnetic bubbles in the vicinity of the active device region.

At 100 and 200 Oe, the propagating mode $m_P$ was excited at small currents, while the localized "bullet" mode $m_L$ appeared at larger currents, similarly to the behaviors at 300 Oe discussed above [Fig.~\ref{fig1}]. An additional peak with two sidebands was observed between the $m_P$ and $m_L$ modes in the intermediate current range [more easily distinguished in the low-temperature data, Fig.~\ref{fig3} below]. Since this peak appeared in the same range of fields as the magnetic bubble domain state, it can be attributed to the dynamical bubble mode $m_B$, as discussed in Ref.~\cite{rhliuPRL2015}. At fields larger than $200$~Oe, the magnetic bubble domain state was destroyed by the out-of-plane component of the  field, and the dynamical bubble mode disappeared.
Mode hopping at currents near $I_t$, and mode coexistence at currents above $I_C$ were observed for all the measurement fields, although the intensity of the propagating mode $m^*_P$ decreased with increasing field, as shown in Fig.~\ref{fig2}(e-l). These behaviors are likely caused by the reduced mode coupling due to the increasing frequency separation between the propagating modes and the localized "bullet" mode, resulting in the gradual suppression of the propagating modes.

According to the theory of multimodal dynamics~\cite{modecoupling,modecoupling2}, mode coupling among the dominant modes plays an essential role in the mode hopping and mode coexistance. The coupling can originate from several different linear and/or non-linear mechanisms, including thermal magnon-mediated scattering arising from the interaction between the dominant modes and thermally excited magnon bath, direct exchange interaction between modes, or their dipolar interaction.

To identify the mode coupling mechanisms in the SHNO, spectroscopic measurements were repeated at a cryogenic temperature $T$ = 6 K, where thermal effects are significantly reduced. Figure~\ref{fig3} shows the dependence of spectral characteristics on $I$, at a representative field H = 700 Oe. The high-frequency propagating mode $m_p$ was observed at small current, followed by an abrupt transition to the low-frequency localized "bullet" mode at $I_t$ = 13 mA, with another high frequency propagating mode appearing near $f_{FMR}$ above $I_c=14$ m A. These behaviors are qualitatively similar to those observed at $T$ =295 K. However, in contrast to the room-temperature data shown in Fig.~\ref{fig2}, the current-dependent spectra of the three dominant modes - high frequency propagating mode $m_P$, intermediate-frequency "bubble skyrmion" soliton mode $m_B$, and the low-frequency localized "bullet" mode $m_L$, observed in certain field ranges [Fig.~\ref{fig4}] - exhibit clearly distinct spectral characteristics, not noticeably affected by the presence of another mode, or proximity to the transition between modes. These behaviors indicate that mode coupling causing mode hopping behaviors near the mode transition in the studied SHNO is strongly suppressed, indicating that thermal magnon-mediated scattering, rather than direct interactions between the modes, is the mechanism of mode coupling in the studied SHNO.

We note that thermal effects also play a significant, but different, role in the magnetization dynamics observed in the absence of applied field [Figs.~\ref{fig2}(a), ~\ref{fig4}(a)]. At $6$~K, the spectrum exhibits a narrow oscillation peak at $2.25$~GHz, with the minimum linewidth $FWHM =$ 5 MHz, without the low-frequency tail observed at $295$~K. Moreover, the presence of a well-pronounced second harmonic of the oscillation clearly demonstrates that a large amplitude of the oscillation is achieved. These results support our interpretation of room-temperature broadening in terms of the interplay between oscillation and thermal fluctuations of the magnetic bubbles, which become suppressed at the cryogenic temperatures.

\begin{figure}[htbp]
\centering
\includegraphics[width=0.4\textwidth]{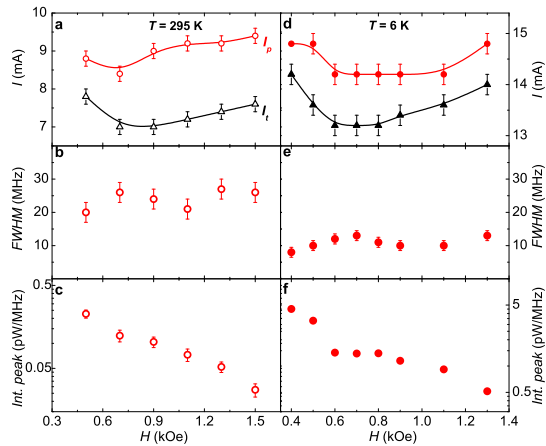}
\caption{Dependence of the microwave generation characteristics on field $H$ at 295 K and 6 K. (a)-(d) The characteristic currents $I_{p}$, and $I_{t}$ defined in the text, (b)-(e) FWHM, (c)-(f) peak intensity at currents $I_{p}$.}\label{fig5}
\end{figure}

To gain further insight into the thermal effects on the generation linewidth of SHNO, we focus on the single-mode excitation regime. We exclude the data obtained at $H$ $\leq$ 300 Oe, since these results are affected by the complex dynamics of bubbles present at small fields. Next, in analyzing the dependence on the magnetic field, we eliminate the possible artifacts coming from the dependence of characteristic currents on field, by focusing on the field-dependent current value $I_{p}$ corresponding to the minimum linewidth, and coinciding with the highest peak power density. As the data in Figs.~\ref{fig5}(a) and (d) show, the current $I_p$ remains far away from the mode transition at $I_t$. Therefore, the spectral characteristics observed at $I_p$ are to a good approximation determined by the direct effects of thermal fluctuations on the single-mode dynamics. The characteristic currents $I_t$ and $I_p$ exhibited a similar overall field dependence, with a broad minimum around $H\simeq$ 700 Oe both at $6$~K and $295$~K [Fig.~\ref{fig5}(a) and (d)], similar to the prior observations for SHNOs with in-plane magnetization~\cite{Demidov_SHO,rhlpra}. While  the maximum peak power generated at $I_p$ exhibited an exponential decrease with increasing field [Fig.~\ref{fig5}(c) and (f)], the minimum linewidth of 25$\pm$5 MHz at room temperature was almost field-independent, and only about 2 times larger than ~10$\pm3$ MHz at 6 K [Fig.~\ref{fig5}(b) and (e)], indicating that the temperature-dependent thermal noise originating from thermal magnons of system is likely the main source for the linewidth broadening~\cite{kim2}.

To summarize, we demonstrated that spin Hall nano-oscillator based on the Pt/[Co/Ni] multilayer with PMA exhibits several distinct dynamical regimes, which can be achieved by varying the applied field and current. The propagating spin wave mode is observed at small currents, transitioning to the localized "bullet" soliton mode, similar to that observed in spin-torque oscillators with in-plane magnetic anisotropy. Besides the propagating and the "bullet" modes, a "bubble skyrmion" soliton mode correlated with the magnetic bubble state at small fields is also observed, enabling coherent auto-oscillations in the absence of the applied field. At room temperature, mode hopping and multimode coexistence, observed over certain current ranges, degrade the spectral characteristics. These effects are remarkably suppressed at a cryogenic temperature, demonstrating that the coupling between dynamical modes is mediated by thermal magnons. Our results suggest a practical approach to controlling the microwave spectral properties of spin Hall nano-oscillators, by tailoring the interactions among the dynamical modes.

L.N.C, Y.W.D and R.H.L are supported by the National Key Research and Development Program of China (2016YFA0300803), National Natural Science Foundation of China (No.11774150), Applied Basic Research Programs of Science and Technology Commission Foundation of Jiangsu Province (BK20170627), and the Open Research Fund of Jiangsu Provincial Key Laboratory for Nanotechnology. S.U. acknowledges support from NSF grant Nos. ECCS-1804198 and DMR-1504449.


\begin{references}
\bibitem{slon1}
J. C. Slonczewski, \Journal{J. Magn. Magn. Mater.} {159}{L1} {1996}.

\bibitem{berger}
L. Berger, \Journal{Phys. Rev. B}{54}{9353}{1996}.


\bibitem{magnonics}
V. V. Kruglyak1, S. O. Demokritov and D. Grundler, J. Phys. D: Appl. Phys. 43 264001(2010).

\bibitem{TorrejonNature2017}
J. Torrejon, M. Riou, F. A. Araujo, S. Tsunegi, G. Khalsa, D. Querlioz, P. Bortolotti, V. Cros, K. Yakushiji, A. Fukushima, H. Kubota, S. Yuasa, M. D. Stiles, and J. Grollier,
 Nature \textbf{547}, 428 (2017).

\bibitem{nonlineartheory1}
A. Slavin, V. Tiberkevich, Physical review letters 95, 237201 (2005).

\bibitem{nonlineartheory2}
J. V. Kim, V. Tiberkevich, A. N. Slavin,  Physical review letters 100, 017207 (2008).


\bibitem{temline}
P. K. Muduli, O. G. Heinonen, J. {\AA}kerman, Physical Review B 86, 174408(2012).

\bibitem{Fert-Tem-Vor}
P. Bortolotti, A. Dussaux, J. Grollier, V. Cros, A. Fukushima, H. Kubota, K. Yakushiji, S. Yuasa, K. Ando, A. Fert, Applied Physics Letters 100, 042408 (2012).


\bibitem{thermaleffect}
J. F. Sierra, M. Quinsat, F. Garcia-Sanchez, U. Ebels, I. Joumard, A. S. Jenkins, B. Dieny, M. C. Cyrille, A. Zeltser, J. A. Katine, Applied Physics Letters 101, 062407 (2012).

\bibitem{SWsCoex}
R. K. Dumas, E. Iacocca, S. Bonetti, S. R. Sani, S. M. Mohseni, A. Eklund, J. Persson, O. Heinonen, J. {\AA}kerman, Physical review letters 110, 257202 (2013).

\bibitem{hoppingmech}
R. Sharma, P. D\"{u}rrenfeld, E. Iacocca, O. G. Heinonen, J. {\AA}kerman, P. K. Muduli, Applied Physics Letters 105, 132404 (2014).

\bibitem{modecoupling}
E. Iacocca, P. D\"{u}rrenfeld, O. Heinonen, J. {\AA}kerman, R. K. Dumas, Physical Review B 91, 104405(2015).


\bibitem{modehoppingtheory}
E. Iacocca, O. Heinonen, P. K. Muduli, J. {\AA}kerman, Physical Review B 89, 054402(2014).

\bibitem{modecoupling2}
S. S. L. Zhang, E. Iacocca, O. Heinonen, Physical Review Applied 8, 014034(2017).


\bibitem{tsoiprl}
M. Tsoi, A.G.M. Jansen, J. Bass, W.C. Chiang, M. Seck, V. Tsoi and P. Wyder, \Journal{Phys. Rev. Lett.}{80}{4281}{1998}.

\bibitem{cornellorig}
J.A. Katine, F.J. Albert, R.A. Buhrman, E.B. Myers, and D.C. Ralph, \Journal{Phys. Rev. Lett.}{84}{3149}{2000}.

\bibitem{miron}
I. M. Miron, G. Gaudin, S. Auffret, B. Rodmacq, A. Schuhl, S. Pizzini, J. Vogel, and P. Gambardella, Nature Materials 9 (3), 230-234 (2010);

\bibitem{liulq}
L. Q. Liu, O. J. Lee, T. J. Gudmundsen, D. C. Ralph, and R. A. Buhrman. \Journal{Phys. Rev. Lett.} {109} {096602} {2012}.

\bibitem{liuprb}
R. H. Liu, W. L. Lim, and S. Urazhdin, \Journal {Phys. Rev. B} {89} {220409(R)}{2014}.

\bibitem{Demidov_SHO}
V.E. Demidov, S. Urazhdin, H. Ulrichs, V. Tiberkevich, A. Slavin, D. Baither, G. Schmitz, and S. O. Demokritov, \Journal{Nat. Mater.} {11}{1028}{2012}.

\bibitem{liuSHNO1}
R.H. Liu, W.L. Lim, and S. Urazhdin, \Journal{Phys. Rev. Lett.} {110}{147601} {2013}

\bibitem{Diakonov}
M. I. Dyakonov and V. I. Perel, \Journal{Sov. Phys. JETP Lett}{13}{467}{1971}.

\bibitem{manchon}
A. Manchon, and S. Zhang. \Journal{Phys. Rev. B.} {79}{094422} {2009}.

\bibitem{rashba}
Y. A. Bychkov and E. I. Rashba, J. Phys. C \textbf{17}, 6039 (1984).
G. Dresselhaus. Phys. Rev. \textbf{100}, 580 (1955).



\bibitem{NSHNO}
V. E. Demidov, S. Urazhdin, A. Zholud, A. V. Sadovnikov, S. O. Demokritov, Applied Physics Letters 105, 172410 (2014).

\bibitem{NWSHNO}
Z. Duan, A. Smith, L. Yang, B. Youngblood, J. Lindner, V. E. Demidov, S. O. Demokritov, I. N. Krivorotov, Nature communications 5, 5616 (2014).

\bibitem{nanoSHNO2}
A. Zholud, S. Urazhdin, Applied Physics Letters 105, 112404 (2014).

\bibitem{secondpeak}
H. Ulrichs, V. E. Demidov, S. O. Demokritov, Applied Physics Letters 104, 042407 (2014)

\bibitem{rhliuPRL2015}
R. H. Liu, W. L. Lim, and S. Urazhdin,  Phys. Rev. Lett. \textbf{114}, 137201 (2015).

\bibitem{mazraati}
H. Mazraati, S. Chung, A. Houshang, M. Dvornik, L. Piazza, F. Qejvanaj, S. Jiang, T. Q. Le, J. Weissenrieder and J. Akerman,
Applied Physics Letters 109, 242402(2016).

\bibitem{mutualSHNO1}
A. A. Awad, P. D\"{u}rrenfeld, A. Houshang, M. Dvornik, E. Iacocca, R. K. Dumas, and J. {\AA}kerman, Nature Physics 13, 292-299 (2016).

\bibitem{mutualSHNO2}
S. Urazhdin, V. E. Demidov, R. Cao, B. Divinskiy, V. Tyberkevych, A. Slavin, A. B. Rinkevich, S. O. Demokritov, Applied Physics Letters 109, 162402 (2016).

\bibitem{PRSHNO}
V. E. Demidov, S. Urazhdin, G. de Loubens, O. Klein, V. Cros, A. Anane, S. O. Demokritov, Physics Reports 673, 1-31 (2017)

\bibitem{SHNOYIG}
V. E. Demidov, M. Evelt, V. Bessonov, S. O. Demokritov, J. L. Prieto, M. Munoz, J. Ben Youssef, V. V. Naletov, G. de Loubens, O. Klein, M. Collet, P. Bortolotti, V. Cros, A. Anane, Scientific reports 6, 32781 (2016).


\bibitem{mosendz}
O. Mosendz, J. E. Pearson, F.Y. Fradin, G.E.W. Bauer, S.D. Bader, and A. Hoffmann, \Journal{Phys. Rev. Lett.} {104}{046601} {2010}.

\bibitem{rhlpra}
R. H. Liu, Lina Chen, S. Urazhdin, and Y.W. Du,
Physical Review Applied 8, 021001 (2017).

\bibitem{kim2}
J.-V. Kim, Phys. Rev. B \textbf{73}, 174412 (2006).


\noindent
\end{references}
\end{document}